\documentclass[ article, aps,]{revtex4-1}
\usepackage{amsmath}

\usepackage{graphicx}
\usepackage{dcolumn}
\usepackage{bm}
\usepackage{color}

\newcommand{\beq}{\begin{equation}}
\newcommand{\enq}{\end{equation}}
\newcommand{\beqa}{\begin{eqnarray}}
\newcommand{\beqast}{\begin{eqnarray*}}
\newcommand{\enqa}{\end{eqnarray}}
\newcommand{\enqast}{\end{eqnarray*}}

\begin{document}

\title{ $\mathrm{pp}$ Interaction at Very High Energies in Cosmic Ray
Experiments }

\author {A. Kendi Kohara , Erasmo Ferreira and Takeshi Kodama}
\address{ Instituto de F\'{\i}sica, Universidade Federal do Rio de
 Janeiro, C.P. 68528, Rio de Janeiro 21945-970, RJ, Brazil}  


\begin{abstract}
 An analysis of  p-air cross section data from  Extensive Air
Shower (EAS) measurements is presented, based on an
analytical representation of the pp scattering amplitudes that describes
with high precision  all available  accelerator data  at ISR, SPS and LHC
energies. The theoretical basis of the representation, together with the 
very smooth energy dependence of parameters controlled by unitarity and dispersion relations, permits reliable  extrapolation to high energy cosmic ray and asymptotic energy ranges. Calculations of $\sigma_{\rm p-air}^{\rm prod}$ based on Glauber formalism are made using the input values of the quantities $\sigma$, $\rho$, $B_I$ and $B_R$ at high energies, with attention given to the independence of the slope parameters, with $ B_{R}\neq B_{I}$ . The influence of contributions of diffractive intermediate states, according to 
Good-Walker formalism, is examined. The comparison with cosmic ray data 
is very satisfactory in the whole pp energy interval from 1 to 100 TeV. High energy asymptotic behavior  of cross sections is  investigated in view of the geometric scaling property  of the amplitudes. The observed energy dependence of the ratio
between p-air and pp cross sections in the data is shown to be related to
the   nature of the pp cross section at high energies, that does not agree with the black disk image.
\end{abstract}



         \maketitle

\section{Introduction   \label{intro}  }

Recently detailed analyses of the experimental pp and $\mathrm{p\bar p}$
scattering data have been performed for the highest energy domain available 
\cite{KEK_2013,KEK_2013b, KEK-to-be}, with determination of amplitudes and
cross sections based on the QCD stochastic vacuum model \cite{ferreira1}.
These analyses lead to very precise quantitative identification of analytic properties  of the imaginary and real parts of the elastic scattering amplitudes, disentangling their presences in the observable
quantities.

The amplitudes are founded on a QCD motivated model \cite{dosch}, controled
by the unitarity and requirements from dispersion relations \cite{ferreira2}%
, thus furnishing a bridge between experimental data and microscopic models.
It has also been shown that the high precision in the description of all
available experimental data covering wide energy domain is attained with
very smooth energy dependence \cite{KEK-to-be}. We have then established a
full $(s,t)$ framework that allows safe interpolations and extrapolations
required in the present era of expansion of the energy frontier. After
successful reproduction of the data in the energy frontier of accelerator
physics at the $\sqrt{s}$ = 7 TeV and 8 TeV energies of LHC, in the present
work we direct our efforts to the examination of the cosmic ray data
extracted from studies of Extensive Air Showers (EAS), where there is access
to pp center of mass energies of up to 100 TeV. We feel that we start to
approach the asymptotic regime where we hope to find the simplified
dynamical description of elastic and diffractive processes in which the
proton enters as a global object, determining the main features of the
observables through its size and the modification of the QCD vacuum around
it. In this high energy regime we may find the ideal conditions for the
application of the concept and method of the Stochastic Vacuum Model \cite%
{dosch} in which our amplitudes are based.

The purpose of the present work is to compare the proton-air production
cross section, calculated in the framework of Glauber model using our
representation of pp scattering as input, to the experimental values
obtained from the available cosmic ray data. We are mainly concerned with
the energies beyond the LHC experiments but also present results for EAS
experiments in the region below 1 TeV.

We also study the behaviour expected for the p-air interaction at ultra-high
energies, both as continuous extrapolation based on the region of the
present data and as consequence of the known properties of the p amplitudes
in $b$-space.

As mentioned above, our proton-proton scattering amplitudes have been
carefully determined, permitting identification of the properties of the
real part which is often neglected in calculations at high energies. We here
stress again the importance of the difference between the slopes $B_{I}$ and 
$B_{R}$ of the imaginary and real parts. In the present work this detail
enters in the application of Glauber formalism to evaluate the connection
between p-air and pp cross sections.

Our analysis of energy dependence of amplitudes and observables in pp
collisions shows that the total cross section has a neat $\log ^{2}{s}$ form 
\cite{KEK-to-be}, as already indicated in several analyses \cite{PDG}. An
important feature of our results is that, the slope parameters, both for $%
B_{I}$ and $B_{R}$, also a $\log ^{2}{s}$ dependence. This is new and
important finding. Generally accepted idea is that the slope of the
differential cross sections varies like simple linear $\log {s}$, as in
Regge phenomenology. Our new result has a crucial effect for the use of
Glauber formalism in the analysis of p-air extended showers at the high
energies of our concern, since the value of the slope $B_{I}$, together with
the value of the total cross section, are the basic and strongly influent
inputs of the calculation.

For the application of Glauber approach, we basically require information on
the amplitudes in forward scattering. In our model these features are easily
obtained taking small $t$ limit \cite{ferreira1,KEK_2013, KEK_2013b} in our
full-$|t|$ treatment. In these conditions the amplitudes take simpler
exponential forms requiring only two parameters to specify each amplitude.
The relevant parameters are then the total cross section $\sigma $, the
ratio $\rho $ between real and imaginary parts at $t=0$, and the slopes $%
B_{I}$ and $B_{R}$ of each of the two parts. Our full-$t$ analysis \cite%
{KEK-to-be} provides the energy dependence of these quantities with simple
analytical forms that are appropriate for the whole energy range from 50 GeV
to 100 TeV. With these forms at hand, we investigate the behaviour of
quantities that are meaningful for the investigation of important features
of the interaction in the forward region, and can make predictions for
asymptotic energies.

The present paper is organized as follows. In the next section, we summarize
our representation and the energy dependence of the necessary parameters for
the application in calculation of p-air cross section in the Glauber
formalism. We also show the high energy asymptotic behavior of quantities
that have finite asymptotic limits, to obtain important information for
extrapolation to the ultra-high energy and asymptotic domains. In 
Sec. \ref{data-section}  
we apply the resuts of the Glauber formalism to calculate p-air cross 
section using our
inputs and compare with the experimental values. We show that the results
for $\sigma _{\mathrm{p-air}}^{\mathrm{prod}}(s)$ can be conveniently put in
simple analytic form with very good accuracy, and then prove that the ratio $%
\sigma _{\mathrm{p-air}}^{\mathrm{prod}}/\sigma _{\mathrm{pp}}$ decreases
slowly, approaching a finite limit at high energies. In Sec. \ref{asymptotic-section} 
 we discuss
the geometric scaling property of our amplitude to understand the asymptotic
behavior of p-air interaction and show how the non-black disk nature of our
pp amplitude affects the asymptotic ratio of pA to pp cross sections. The
last section is devoted to summary and discussion of the present work.

\section{Forward Scattering Amplitudes  \label{forward}}

In the treatment of elastic pp and p$\mathrm{{\bar{p}}}$ scattering in the
forward direction, with amplitudes approximated by pure exponential forms,
the differential cross section is written 
\begin{eqnarray}
\frac{d\sigma }{dt} &=&\pi \left( \hbar c\right) ^{2}~~\Big\{\Big[\frac{\rho
\sigma }{4\pi \left( \hbar c\right) ^{2}}~{{e}^{B_{R}t/2}+F^{C}(t)\cos {%
(\alpha \Phi )}\Big]^{2}}  \nonumber \\
&&+\Big[\frac{\sigma }{4\pi \left( \hbar c\right) ^{2}}~{{e}%
^{B_{I}t/2}+F^{C}(t)\sin {(\alpha \Phi )}\Big]^{2}\Big\}~,}
\label{diffcross_eq}
\end{eqnarray}%
where $t\equiv -|t|$ and we must allow different values for the slopes $B_{I}
$ and $B_{R}$ of the imaginary and real amplitudes. With $\sigma$ in
milibarns and $|t|$ in GeV$^2$, we have $\left( \hbar c\right)^2~= ~0.3894$.
Since we work with $B_{R}\neq B_{I}$ , treatment of the Coulomb interference
requires a more general expression for the Coulomb phase, which has been
developed before \cite{KEK_2013}. However, in the present work we only need
the forward ($|t|=0$ ) nuclear amplitudes and slopes, and the Coulomb
interaction does not enter, so that we put $F^C(t)=0$.

\begin{figure}[b]
\includegraphics[width=8cm]{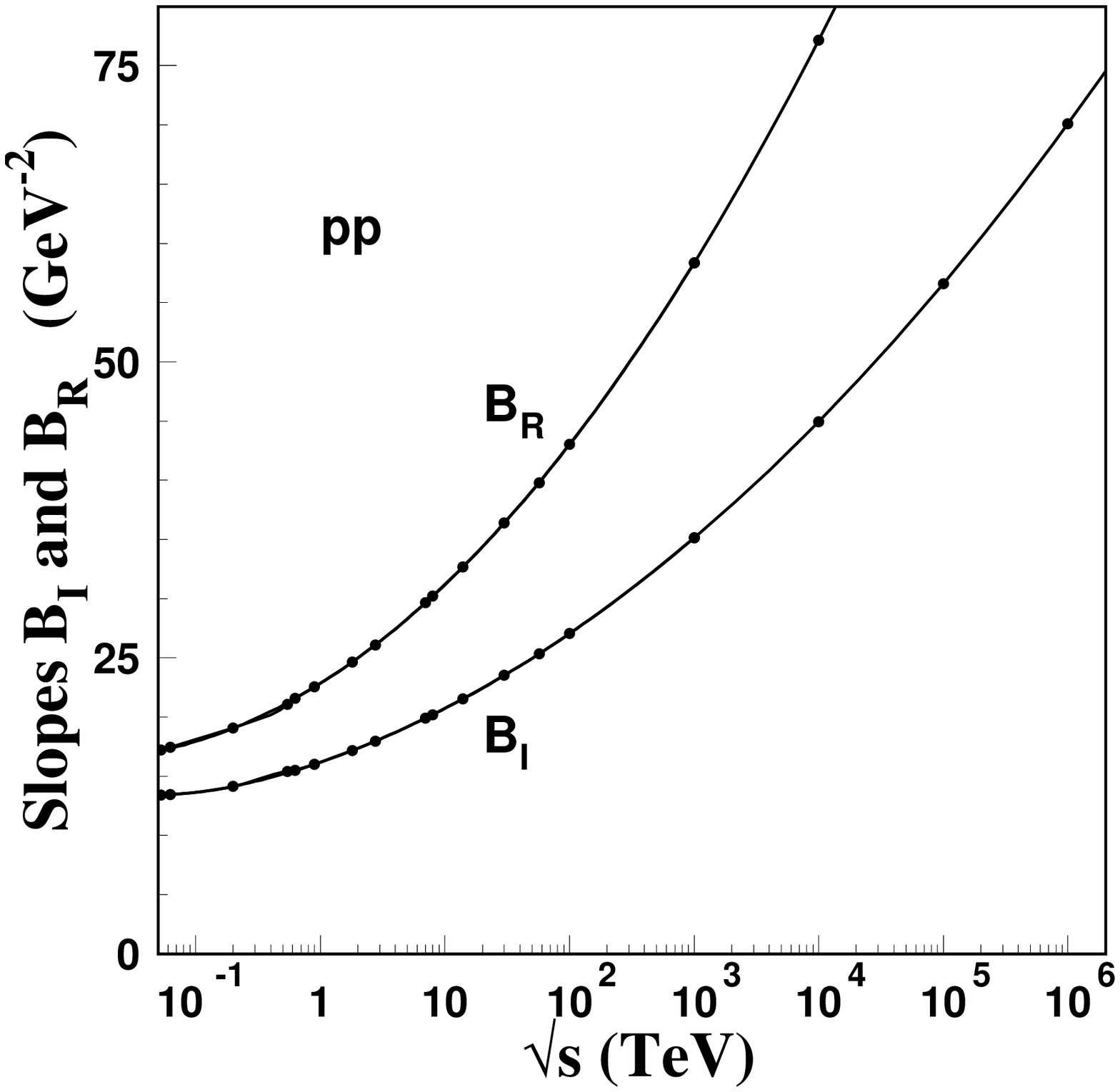}  %
\includegraphics[width=8cm]{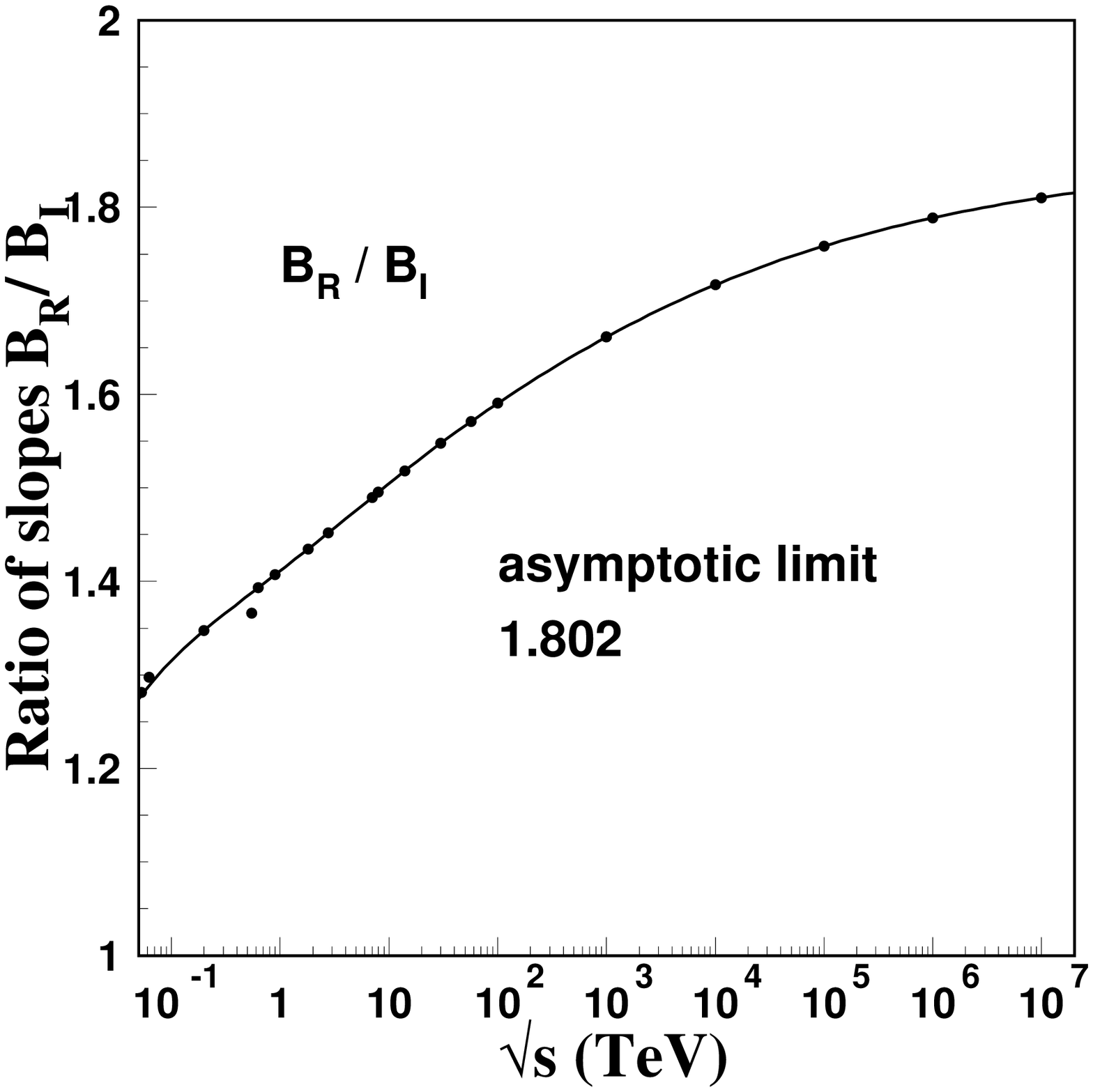} 
\caption{ The slopes of the imaginary and real parts of the amplitude
increase with the energy as $\log^2 \protect\sqrt{s}$, always with $B_R~ > ~
B_I$. The asymptotic value of the ratio is 1.802 . }
\label{slopesfig}
\end{figure}

The energy dependences of the four quantities are given by 
\begin{equation}  \label{sig-eq}
\sigma(s)= 69.3286+12.6800\log\sqrt{s}+1.2273\log^2 \sqrt{s} ~,
\end{equation}
\begin{equation}  \label{BI-eq}
B_I(s) = 15.7848 + 1.75795 \log\sqrt{s}+0.149067\log^2\sqrt{s} ~ ,
\end{equation}
\begin{equation}  \label{BR-eq}
B_R(s) = 22.8365 + 2.86093 \log{\sqrt{s} }+0.329886 \log^2\sqrt{s} ~,
\end{equation}
and 
\begin{equation}  \label{rho-eq}
\rho(s)= \frac{3.528018+0.7856088 \log\sqrt{s} } {25.11358+4.59321 \log\sqrt{%
s}+0.444594\log^2 \sqrt{s} } ~ ,
\end{equation}
where $\sqrt{s}$ is in TeV, $\sigma$ in milibarns, $B_I$ and $B_R$ are in $%
\nobreak\,\mbox{GeV}^{-2}$; $\rho$ is dimensionless, passes through a
maximum at about 1.8 TeV, and decreases at higher energies, with asymptotic
value zero. The ratio $B_R/B_I$ is always larger than one, as expected from
dispersion relations \cite{ferreira2}, and behaves asymptotically like 
\begin{equation}
\frac{B_R}{B_I} \rightarrow 1.80198 + \frac{4.82272}{\log\sqrt{s}} - \frac{%
118.192} {\log^2\sqrt{s} } ~.
\end{equation}
This ratio is not a monotonic function, having a small bump (it goes up to
1.86) at very large energies with $\log \sqrt{s} \approx 30 - 40$ ,  and
then decreases towards its asymptotic limit. The slopes and their ratio are
shown in Fig. \ref{slopesfig}. 

The dimensionless ratio 
\begin{equation}  \label{ratioRI}
R_I = \frac{1}{ (\hbar c)^2 } \frac {\sigma}{16\pi B_I}
\end{equation}
is often studied in considerations about the form of the pp interaction. The
factor $(\hbar c)^2 $ is included to allow practical use of mixed units for $%
\sigma$ (usually in milibarns) and $B_I$ (usually in $\nobreak\,\mbox{GeV}%
^{-2}$). In our description of the pp system, as given by the energy
dependences in Eqs. (\ref{sig-eq}-\ref{rho-eq}), this ratio has the high
energy behaviour 
\begin{equation}  \label{ratioRIasymp}
R_I =\frac{1}{ (\hbar c)^2 } \frac {\sigma }{16\pi B_I} \rightarrow 0.341775+%
\frac{1.50046}{\log\sqrt{s}}-\frac{30.2842}{\log^2\sqrt{s}} ~ .
\end{equation}
This quantity is not monotonically varying, passing through a small bump in
a range at very large energies, and then moving towards the asymptotic limit
0.342 .

For amplitudes of pure exponential behaviour, as we have in this paper, this
ratio is numerically equal to the ratio $\sigma _{\rm pp}^{\rm{el,I}}/\sigma $
between integrated elastic and total pp cross section. Thus this elastic
ratio is also nearly 1/3, and the inelastic ratio is $\sigma _{\mathrm{pp}}^{%
\mathrm{inel}}/\sigma ~\approx ~2/3$. We thus observe that the ratio is far
from the value 1/2 that is characteristic of the idea of a black disk, where
the interaction, considered as function of the impact parameter, is maximal
inside a range $b_{0}$ and zero outside this range. The conjecture of some
authors is that at infinite energy the pp interaction could take the form of
a black disk, as consequence of a kind of geometric scale property. Our
results show that there is no such black disk behaviour. In our case, we
observe an approximate geometrical scaling in the $b$-space differential
cross sections $d^{2}\sigma ^{\mathrm{tot}}/d^{2}\vec{b}$ that start nearly
constant (equal to 2) , and then decrease in a scaled way, forming a
diffused surface region. For the black disk instead the cross section
behaves as the Heaviside step function. We show that the diffused range at
high energies is responsible for the values of the ratios 
$\sigma_{\rm pp }^{\rm el,I}/\sigma $   and  
$\sigma_{\rm pp}^{\rm inel}/\sigma $ that are asymptotically different 
from 1/2. Details are presented
and discussed in Sec. \ref {asymptotic-section}.

We remark that we have used the slope $B_I$ in the ratio (\ref{ratioRI})
defined above. We may similarly define the ratio using the $B_R$ slope, and
then we obtain the high $\sqrt{s}$ behaviour 
\begin{equation}  \label{ratioRR}
R_R=\frac{1}{ (\hbar c)^2 }\frac { \sigma}{16\pi B_R} \rightarrow 0.1896 + 
\frac{0.325061}{\log\sqrt{s}} - \frac{5.23579}{\log^2{\sqrt{s} } } ~.
\end{equation}
With pure exponential form in the real amplitude, this fraction is equal to
the ratio $( \sigma_{\rm pp}^{\rm el,R}/{\rho^2}) / \sigma $ . Since $\rho$ is
small, the contribution of the real part to the integrated elastic cross
section is also small.

\begin{figure}[b]
\includegraphics[width=8cm]{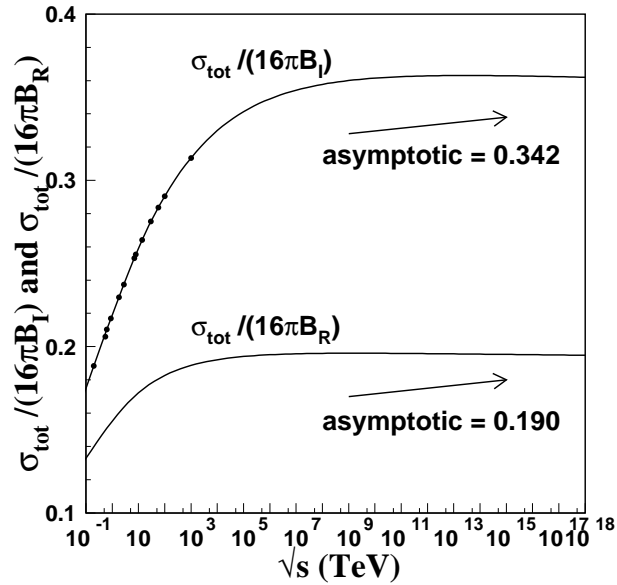}   
\caption{ Energy dependence of the dimensionless ratios between total pp
cross section and the slopes $B_I$ and $B_R$, as defined by Eqs. (\protect
\ref{ratioRI}, \protect\ref{ratioRR}). The expressions have finite
asymptotic limits, as shown in equations and in the plots.}
\label{trueratios} 
\end{figure}

The energy dependence of the two ratios $R_I$ and $R_R$ is shown in 
Fig. \ref{trueratios}.

\section{ Glauber calculation  \label{Glauber-section}}

 The information on the parameters given above for the pp interaction enters
in the calculation of production cross section $\sigma_{\mathrm{p-air}}^{%
\mathrm{prod}} $ that is obtained from the analysis of Extensive Air Showers.

Glauber method \cite{Glauber} provides the basic principles for the
calculation of strong interactions with composite systems. The method first
introduced in the treatment of scattering by deuterons was extended to more
general nuclei, where the complexity of rescattering processes lead to
considerations about the importance of intermediate diffracted states \cite%
{Good_Walker} not given as known external inputs. The application of the
method to the analysis of proton-air collisions in the Extensive Air Showers
(EAS) \cite{Engel}  gives the basic connection between the cosmic ray data
and the hadronic scattering properties. As the basis of Glauber formalism is
well know in its standard form , we present here the essential points giving
the connection between pp and p-air processes, emphasizing the new features
that arise from our treatment of pp amplitudes.

Our forward amplitudes $(s,t)$ show different $t$ behaviour in the imaginary
and real parts, with different slopes $B_I$ and $B_R$. Transferred to b
space, we write amplitudes 
\begin{eqnarray}
\widehat T_{\mathrm{pp}}(s,\vec b)&=& \widehat T_R(s,\vec b)+i\widehat
T_I(s,\vec b) \\
&=& \frac{\sigma^{\mathrm{tot}}_{\mathrm{pp}}} {4\pi(\hbar c)^2}\bigg[ \frac{%
\rho}{B_R} e^{-\frac{b^2}{2 B_R}}+i \frac{1}{B_I} e^{-\frac{b^2}{2 B_I}} %
\bigg] ~.  \nonumber
\end{eqnarray}
In terms of the eikonal function $\chi(s,\vec b)$ this is written  
\begin{equation}  \label{iT-eikonal}
-i ~\widehat T_{\mathrm{pp}}(s,\vec b)~ = ~ 1 - e^{i \chi_{\mathrm{pp}%
}(s,\vec b)} ~ \equiv ~ \Gamma_{\mathrm{pp}}(s,\vec b) ~.
\end{equation}
The term $e^{i \chi_{\mathrm{pp}}(s,\vec b)}$ represents the S-matrix
function in $b$-space. The optical theorem for pp scattering appears as 
\begin{equation}  \label{optical_pp}
\sigma^{\mathrm{tot}}_{\mathrm{pp}}(s) ~ = ~2 ~(\hbar c)^2 ~\Re~\int d^2\vec
b ~ \Gamma_{\mathrm{pp}}(s,\vec b) ~ .
\end{equation}

Analogously, for elastic scattering in the p-A system, we define a
quantity $\Gamma_{\mathrm{pA}}(s,\vec b) $ that satisfies the optical
theorem for the pA total cross section 
\begin{equation}  \label{Sigma_total_H_Ar_2}
\sigma_{\mathrm{pA}}^{\mathrm{tot}} (s) ~ = ~ = ~2 ~(\hbar c)^2 ~\Re~\int
d^2\vec b ~ \Gamma_{\mathrm{pA}}(s, \vec b) ~ .
\end{equation}
Glauber theory introduces a structure to express $\Gamma_{\mathrm{pA}%
}(s,\vec b) $ in terms of pp scattering amplitudes and reaction matrix
elements.

To describe the phenomena in the Extensive Air Showers (EAS) in Cosmic Ray
(CR) observations we need to evaluate the quantity 
\begin{eqnarray}  \label{sigma_producao}
\sigma_{\mathrm{p-air}}^{\mathrm{prod}}=\sigma^{\mathrm{tot}}_{\mathrm{p-air}%
}-(\sigma_{\mathrm{p-air}}^{\mathrm{el}}+\sigma_{\mathrm{p-air}}^{\mathrm{%
q-el}})~
\end{eqnarray}
that is determined experimentally. The quantities named p-air are averages 
over a mixture of nitrogen and oxygen nuclei. 
 
For elastic and quasi-elastic processes characterized by momentum transfer $%
|t|$, a transition matrix element between states $i$ and $f$, defined with
nucleon coordinates $(\vec{r_{1}},...,\vec{r_{A}})$ is written 
\begin{eqnarray}
&&-i ~T_{\mathrm{pA}}^{fi}(s,q^{2})=\frac{1}{2\pi }\int d^{2}\vec{b}~e^{ic%
\vec{q}.\vec{b}}~\int \psi _{f}^{\ast }(\vec{r_{1}},...,\vec{r_{A}}) 
\nonumber \\
&&\times \Gamma _{\mathrm{pA}}(s,\vec{b},\vec{s}_{1},...,\vec{s}%
_{A})~\psi _{i}(\vec{r_{1}},...,\vec{r_{A}})\prod_{j=1}^{A}~d^{3}\vec{r_{j}}%
~,
\end{eqnarray}%
with $\vec{b}$ the p-A impact parameter, $\vec{r}_{i}$ the position of the
nucleon inside the nucleus, $\vec{s}_{i}$ the projection of $\vec{r}_{i}$ in
the perpendicular collision plane.

Glauber method introduces for p-A scattering the expression based on product
of S-matrix factors of $A$ independent elementary scattering processes 
\begin{equation}
\Gamma _{\mathrm{pA}}(s,\vec{b},\vec{s}_{1},...,\vec{s}%
_{A})=1-\prod_{j=1}^{A}\Big[1-\Gamma _{\mathrm{pp}}(s,|\vec{b}-\vec{s_{j}}|)%
\Big]~.  \label{Gamma_p-air-1}
\end{equation}%
This is an assumption of a factorization property for the p-A system. 

Then the expression for the transition matrix element becomes 
\begin{eqnarray}
&&T_{\mathrm{pA}}^{fi}(s,q^{2})=\frac{1}{2\pi }\int d^{2}\vec{b}~e^{ic%
\vec{q}.\vec{b}}\int ~\psi _{f}^{\ast }(\vec{r_{1}},...,\vec{r_{A}})\times  
 \\
&&\bigg[1-\prod_{j=1}^{A}~\Big[1-\Gamma _{\mathrm{pp}}(s,|\vec{b}-\vec{s_{j}}%
|)\Big]\bigg]~\psi _{i}(\vec{r_{1}},...,\vec{r_{A}})\prod_{j=1}^{A}~d^{3}%
\vec{r_{j}}~   ~ . \nonumber \\
\end{eqnarray}

The sum of elastic and quasi-elastic processes is given by 
\begin{eqnarray}  \label{quasi}
&& \sigma_{\mathrm{pA}}^{\mathrm{el}}+\sigma_{\mathrm{pA}}^{\mathrm{%
q-el}}=(\hbar c)^2\int ~d^2\vec q ~ \sum_f |T_{\mathrm{pA}}^{fi}(s,q^2)|^2  \\
&&= (\hbar c)^2\int ~d^2\vec q ~ \sum_f \Bigg|\frac{1}{2\pi}\int d^2\vec b~
e^{ic\vec{q}.\vec{b}}\int ~\psi_{f}^{*}(\vec {r_1},...,\vec {r_A})\times 
\nonumber \\
&& \bigg[ 1- \prod_{j=1}^{A}~\Big[1-\Gamma_{\mathrm{pp}}(s,|\vec b-\vec{s_j}%
|)\Big] \bigg] ~\psi_{i}(\vec {r_1},...,\vec {r_A}) \prod_{k=1}^A~d^3\vec{r_k%
} ~ \Bigg|^2  \nonumber \\
&&=(\hbar c)^2\int d^2\vec b ~ \times   \nonumber \\
&&\int\Bigg|1-\prod_{j=1}^{A}\bigg[1-\Gamma_{\mathrm{pp}}(s,|\vec b-\vec{s_j}%
|)\bigg]\Bigg|^2 \prod_{k=1}^A~\rho_k(\vec{r_k})~d^3\vec{r_k} ~  \nonumber \\ 
&& \equiv (\hbar c)^2\int d^2\vec b  ~   \frac{d\tilde\sigma_{\rm pA}^{\rm{el+q-el} }}{d^{2}\vec{b}} (s,b) ~ . \nonumber   %
 \end{eqnarray}

In  Eq. (\ref{quasi}) we have made use  of the orthogonality condition 
\begin{eqnarray}  \label{orthogonality}
\int\psi_{f}^{*}(\vec {r_1},...,\vec {r_A})\psi_{i}(\vec {r_1},...,\vec {r_A}%
)\prod_{j=1}^{A}d^3\vec{r_j}=0 ~,
\end{eqnarray}
the completeness relation 
\begin{eqnarray}  \label{completeness}
\sum_f^A\psi_{f}^{*}(\vec {r_1},...,\vec {r_A})\psi_{f}(\vec {r_1}^{^{\prime
}},...,\vec {r_A}^{^{\prime }})=\prod_{j=1}^{A}\delta(\vec {r_j}-\vec {r_j}%
^{^{\prime }}) ~
\end{eqnarray}
and the definition of the nucleon densities $\rho_k(\vec r_k)$.

Assuming that the $i$ and $f$ states are similar bound nuclei with nucleon
densities $\rho_j(\vec r_j) $ , and that there is no correlation between the
nucleons in the collision process, we write 
\begin{equation}  \label{density}
\psi_{i}^{*}(\vec {r_1},...,\vec {r_A})\psi_{i}(\vec {r_1},...,\vec {r_A})~
= \prod_{j=1}^{A}\rho_j(\vec r_j) ~,
\end{equation}
where $\rho_j(\vec{r_j})$ is the density of the nucleon $j$ in the nucleus.

For atoms with atomic numbers $A$ less than or equal 18 typically present in
the atmosphere the nuclear densities can be described by harmonic potentials
with s and p orbitals $\rho_s(\vec b)$ and $\rho_p(\vec b)$ that are
introduced explicitly \cite{nucleus} as 
\begin{eqnarray}  \label{shells}
\rho_s(\vec r)&=& \frac{1}{\pi^{3/2}b_0^3}e^{-r^2/b_0^2}~ \\
\rho_p(\vec r)&=& \frac{2r^2}{3\pi^{3/2}b_0^5}e^{-r^2/b_0^2}~,  \nonumber
\end{eqnarray}
normalized to unity 
\begin{equation}  \label{norma}
\int d^3\vec r~\rho_{s,p}(\vec r)=1~ .
\end{equation}
In this work, for nitrogen and oxygen nuclei the parameters are $%
b_0=1.7069~\mathrm{fm}$ and $b_0=1.8133~\mathrm{fm}$ respectively.



Taking the product of Eq.(\ref{quasi}) over the nuclear densities, with 4
nucleons in s shell and A-4 in p shell, we have 
\begin{eqnarray}  \label{sigma_el_qel_4}
&&\sigma_{\mathrm{pA}}^{\mathrm{el}}+\sigma_{\mathrm{pA}}^{\mathrm{q-el%
}}=(\hbar c)^2\int d^2\vec b~\times \\
&&\Bigg\{1-2\Re~\Bigg[\Big[\int d^3\vec{r}\Big(1-\Gamma_{\mathrm{pp}}(\vec b-%
\vec{s})\Big)\rho_s(r)\Big]^4~\times  \nonumber \\
&& \Big[\int d^3\vec{r}\Big(1-\Gamma_{\mathrm{pp}}(\vec b-\vec{s})\Big)%
\rho_p(r)\Big]^{A-4}\Bigg]  \nonumber \\
&+&\Big[\int d^3\vec{r}\Big(1-2\Re~ \Gamma_{\mathrm{pp}}(\vec b-\vec{s}) +
|\Gamma_{\mathrm{pp}}(\vec b-\vec{s})|^2\Big)\rho_s(\vec{r})\Big]^4~\times 
\nonumber \\
&&\Big[\int d^3\vec{r}\Big(1-2\Re~ \Gamma_{\mathrm{pp}}(\vec b-\vec{s}) +
|\Gamma_{\mathrm{pp}}(\vec b-\vec{s})|^2\Big)\rho_p(\vec{r})\Big]^{A-4}%
\Bigg\}~.  \nonumber
\end{eqnarray}

The quantity that enters Eq. (\ref{Sigma_total_H_Ar_2}) for the evaluation
of the total pA cross section is 
\begin{eqnarray}  \label{Gamma_p-air}
&& \Gamma_{\mathrm{pA }}(s,\vec b) \\
&& = 1- \prod_{j=1}^{A}\int d^3\vec{r_j}~\rho_j(\vec{r_j})~\Big[1-\Gamma_{%
\mathrm{pp}}(s,|\vec b-\vec{s_j}|)\Big] ~ .  \nonumber
\end{eqnarray}
and the pA total cross section is given by 
\begin{eqnarray}  \label{optical_2}
&&\sigma_{\mathrm{pA}}^{\mathrm{tot}} (s) ~ = ~2 ~(\hbar c)^2 ~\Re~\int
d^2\vec b\times ~ \\
&& \Bigg(1- \prod_{j=1}^{A}\int d^3\vec{r_j}~\rho_j(\vec{r_j})~\Big[%
1-\Gamma_{\mathrm{pp}}(s,|\vec b-\vec{s_j}|)\Big]\Bigg)   \nonumber \\
&&\equiv (\hbar c)^2 \int d^2\vec b ~ 
 \frac{d\tilde\sigma_{\rm pA}^{\rm tot}}{d^{2}\vec{b}}  (s,b) ~ . \nonumber
\end{eqnarray}

For p-A elastic scattering   we have 
\begin{eqnarray}
\label{elastic_only}
&& \sigma_{\rm pA}^{\rm el}(s)=(\hbar c)^2\int |T_{\rm pA}^{ii}(s,q^2)|^2~d^2\vec q    \\ 
&&= (\hbar c)^2\int \Big|\frac{1}{2\pi}\int d^2\vec b~ e^{ic\vec{q}.\vec{b}}\int ~\psi_{i}^{*}(\vec {r_1},...,\vec {r_A})\times  \nonumber \\ 
&&  \bigg[  1- \prod_{j=1}^{A}~\Big[1-\Gamma_{\rm{pp}}(s,|\vec b-\vec{s_j}|)\Big] \bigg] ~\psi_{i}(\vec {r_1},...,\vec {r_A})    
\prod_{j=1}^A~d^3\vec{r_j}  ~ \Big|^2~d^2\vec q    \nonumber \\  
&&=(\hbar c)^2\int d^2\vec b~\Big| 1- \prod_{j=1}^{A}\int d^3\vec{r_j}~\rho_j(\vec{r_j})~\Big[1-\Gamma_{\rm{pp}}(s,|\vec b-\vec{s_j}|)\Big]\Big|^2 ~  \nonumber\\
&&=(\hbar c)^2\int d^2\vec b~\Big| \Gamma_{\rm pA }(s,\vec b)\Big|^2 ~    
  \equiv (\hbar c)^2 \int d^2\vec b ~ 
 \frac{d\tilde\sigma_{\rm p-air}^{\rm el}}{d^{2}\vec{b}}  (s,b)  ~ .  \nonumber
\end{eqnarray}

We thus follow Glauber formalism \cite{Glauber} in general lines, with 
independent slopes  $B_{R}$ and $B_{I}$. We consider also the effect of the
contributions of intermediate diffractive states according to Good-Walker 
\cite{Good_Walker}, with a parameter $\lambda $. For practical
implementation \cite{Engel} we   re-write Eq.(\ref{Gamma_p-air}) as 
\begin{eqnarray}
&&\Gamma _{\mathrm{pA}}(s,\vec{b},\vec{s}_{1},...,\vec{s}_{A})=1-\frac{1}{%
2}\prod_{j=1}^{A}~\Big[1-(1+\lambda )\Gamma _{\mathrm{pp}}(\vec{b}-\vec{s}%
_{j})\Big]  \nonumber \\
&&-\frac{1}{2}\prod_{j=1}^{A}\Big[1-(1-\lambda )\Gamma _{\mathrm{pp}}(\vec{b}%
-\vec{s}_{j})\Big]~,  \label{diffraction5}
\end{eqnarray}%
and consequently modify Eqs.(\ref{Sigma_total_H_Ar_2}) and (\ref%
{sigma_el_qel_4}).

Stressing that we provide reliable information on cross sections and
amplitude slopes for the pp scattering input, and a proper, although simple,
treatment of Glauber framework, we believe that our calculations of $\sigma_{%
\mathrm{p-air}}^{\mathrm{prod}}$ are worth as a study of the EAS data.
Actually, we show in the next section that there is very good coherence
between our calculations and the data.

\begin{figure}[b]
 \includegraphics[width=8cm]{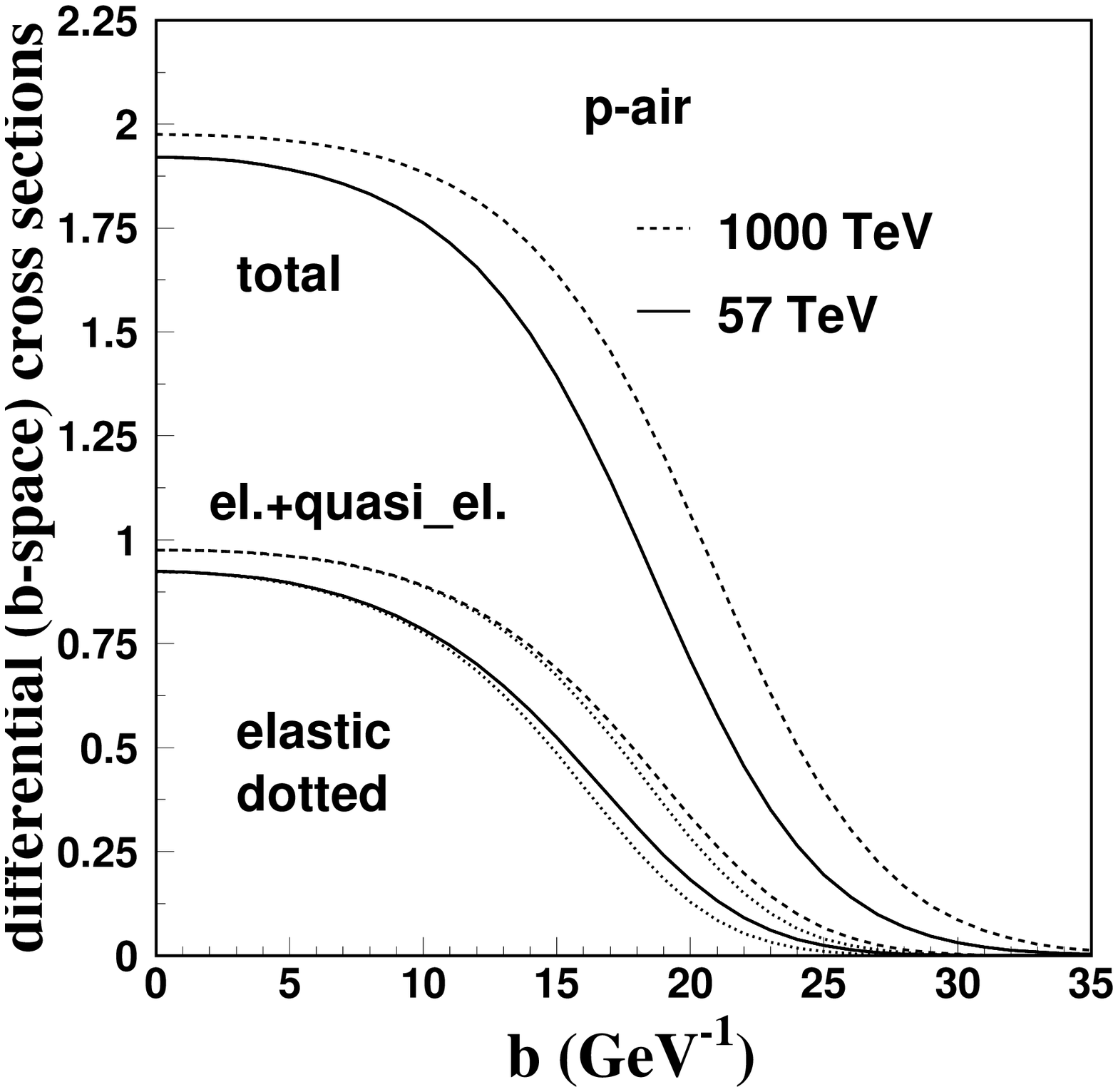} 
 \includegraphics[width=8cm]{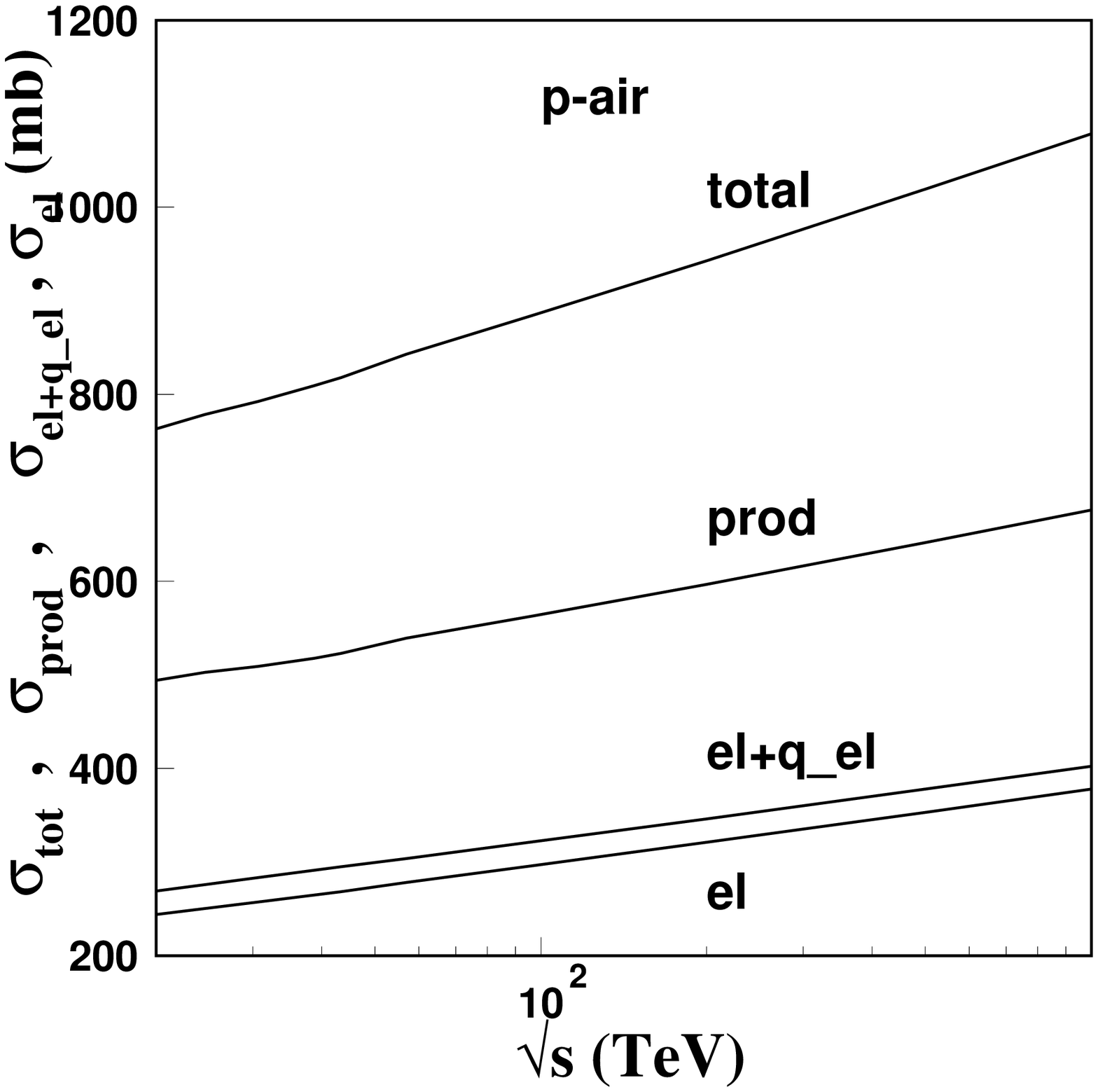} 
\caption{ The quantities 
    $   d\sigma_{\rm p-air}^{\mathrm{tot} }/{d^{2}\vec{b}}   $ 
and $  d\sigma_{\rm p-air}^{\mathrm{el+q-el} }/{d^{2}\vec{b}}    $ 
are plotted as functions of the p-air impact parameter $b$ for the energies  57 
and 1000 TeV. As the energy increases, the saturation limits  2 and 1 are 
approached by the total and inelastic parts for small  $b$. 
The integrated quantities are shown in  the second part of the figure. 
The small difference between elastic+quasi-elastic and purely elastic 
terms is remarkable. }
  \label{p_air-fig}  
 \end{figure}

The  dimensionless quantities that give the $b$-dependence of the
total, elastic+quasi-elastic and pure elastic cross sections  
for the p-air system (taking averages over nitrogen   and oxygen 
components) 
\begin{equation}
\frac{d\tilde\sigma_{\rm p-air}^{\rm tot}}{d^{2}\vec{b}} (s,b) ~ ,  
   \frac{d\tilde\sigma_{\rm p-air}^{\rm{el+q-el} }}{d^{2}\vec{b}}(s,b) ~ ,  
   \frac{d\tilde\sigma_{\rm p-air}^{\rm el}}{d^{2}\vec{b}}  (s,b)  
 \end{equation}
are represented in Fig. \ref{p_air-fig} for the energies  $\sqrt{s}=57$ 
and $\sqrt{s}=1000 $ TeV.  As in the pp system, the total and inelastic 
cross sections for small $b$ approach the limits 2 and 1 as the energy 
increases.  There is little  difference  between the elastic+quasi-elastic 
and the pure elastic quantities.

The integrated quantities $ \sigma_{\mathrm{p-air}}^{\mathrm{tot}} (s),  
 \sigma_{\mathrm{p-air}}^{\mathrm{el}}+\sigma_{\mathrm{p-air}}^{\mathrm{q-el }} (s) ~  $
and  $  \sigma_{\rm p-air}^{\rm el} (s) $
are shown in the second part of the same figure. The ratio 
$\sigma_{\rm p-air}^{\rm el}/\sigma_{\rm p-air}^{\rm tot}$ is 0.33 at 
57 TeV and 0.35 at 1000 TeV. The difference between elastic+quasi-elastic and 
purely elastic contributions is remarkably small, of about 18 \% at 50 GeV and 
falling steadily to zero as the energy increases. The inelastic 
p-air cross section is about 2/3 of the total, as in the pp system.  

\section{ Comparison with Data     \label{data-section} }

\begin{figure}[b]
\includegraphics[width=8cm]{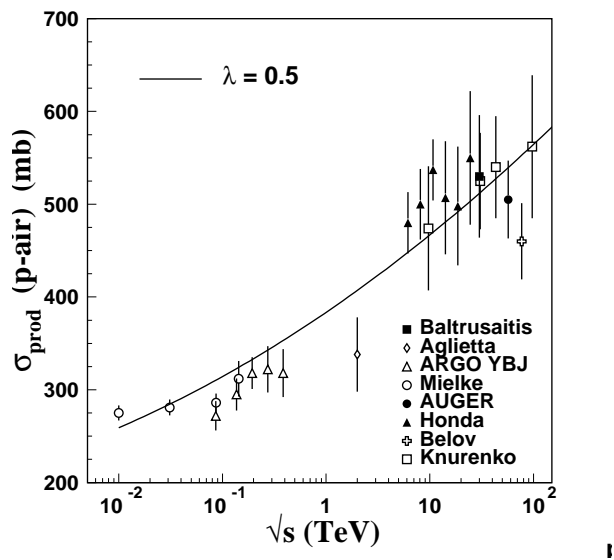}  
\caption{ Our calculation of the p-air production cross section is
represented by the solid line, that is well represented by Eq. (\protect\ref{curve_data}).
Details are given in the text. The data are from several experiments 
\protect\cite{Auger,Belov,Baltru,Honda, Knurenko,Aielli,Mielke,Aglietta}.
Both data and calculations increase with the energy with a $\log^2 \protect%
\sqrt{s}$ form. }
\label{CR_data-fig}  
\end{figure}

Fig. \ref{CR_data-fig} shows our calculation of $\sigma_{\mathrm{p-air}}^{%
\mathrm{prod}}$ with a solid line, together with the data points from
experiments with Extensive Air Showers  \cite{Auger,Belov,Baltru,Honda,
Knurenko,Aielli,Mielke,Aglietta} . 

The procedure is  straightforward and unique, without free parameters, 
made with inputs given by our model for the pp interaction that 
describes the    elastic
differential cross sections at all energies from 20 GeV to 8 TeV
in the whole $t$-range, with high precision. 
For the application in Glauber calculation
of the p-air processes in the EAS experiments, the model enters only in its 
forward scattering limit, and is represented  
by Eqs. (\ref{sig-eq}-\ref{rho-eq}). 
The log-squared increases of $\sigma$, $B_I$, $B_R$  are consequence of 
the Yukawa-like behaviour of the amplitudes, and do not violate unitarity or 
dispersion relations \cite{KEK-to-be}. 
Thus we consider that this is a reliable input.

The calculation of $\sigma_{\mathrm{p-air}}^{\mathrm{prod}}$   
 is made with Eq. (\ref{sigma_producao}),   as explained 
in the previous section. 
 The figure shows that  
in general there is good agreement, without systematic deviation that 
could require additional term  in Eq. (\ref{sigma_producao}) that
would  be  beyond the basic Glauber form. At high energies above 
10 TeV  ($\sqrt{s}$ in the proton-proton system)   the agreement is 
particularly satisfactory,   considering the  quality of the present 
experimental  information. In the low energy region we observe  that 
data from the ARGO-YBJ experiment \cite{Aielli} is  below the theoretical curve,
while   the data from the  Kaskade experiment \cite{Mielke} do not 
shown the same systematic deviation. 
  
The theoretical curve for the production cross section
can be put in the simple and convenient form 
\begin{equation}  \label{curve_data}
\sigma_{\mathrm{p-air}}^{\mathrm{prod}}=383.474+33.158 \log \sqrt{s}+1.3363
\log^2\sqrt{s} ~ ,
\end{equation}
with $\sqrt{s}$ in TeV.  

\begin{figure}[b]
\includegraphics[width=8cm]{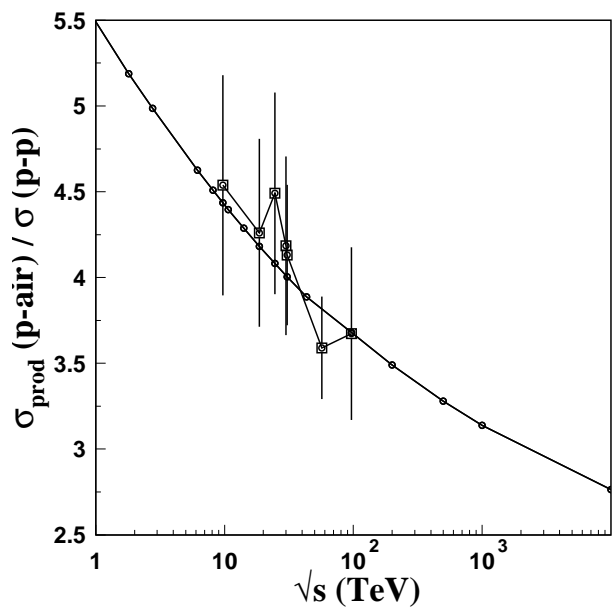} 
\caption{ Ratio of p-air and pp cross sections. We show our calculation in
solid line (with dots) together with selected data. We observe regular behaviour in the
energy variation of the data, that slowly approaches a finite
asymptotic limit.}
 \label{ratio_sigmas-fig}  
\end{figure}

We observe that the data and our calculations of $\sigma_{\mathrm{p-air}}^{%
\mathrm{prod}}$ increase with similar  $\log^2 \sqrt{s} $  energy dependence
as the pp cross sections, but more slowly.   
To compare the two rates  and   give more evidence of
regularity in the data, we show in 
Fig. \ref{ratio_sigmas-fig} the relation 
$\sigma_{\mathrm{p-air}}^{\mathrm{prod}}/\sigma(\rm pp) $ for a set of selected 
data (chosen by regularity reasons) together with our calculations.
The ratio decreases regularly, approaching a finite and distant asymptotic limit,
as pointed out by the relation of forms in  Eqs. (\ref{curve_data}) and (\ref{sig-eq}). 
The importance of the existence of a finite asymptotic limit for this ratio and
its numerical value at ultra-high energies are discussed in a geometric approach 
in Sec.5.
 
We hope that this observation of regularity and interesting energy
dependence of this ratio will be confirmed by more measurements 
and will help  the understanding of the hadronic interactions in 
cosmic ray experiments.

 Other models of the  pp interaction     \cite{Gaisser} 
have different features, such as the energy  dependence of the slopes 
and their  correlation with the   total cross section, and  
the behaviour of the inelastic pp cross section (in our model we have
at high energies $\sigma^{\rm inel}/\sigma^{\rm tot}=2/3$ while the black disk
value is 1/2). 
  The use of these models as pp  inputs may lead 
to systematic deviations with respect to data, and may lead   to  
suggestions of additional contributions 
to the quantity  $\sigma_{\rm p-air}^{\rm prod}$ written  
in Eq. (\ref{sigma_producao}).  
 Thus, as a historical example, the data of Akeno \cite{Honda}  and Fly's Eye \cite{Baltru} 
in the 30 TeV region was studied critically  \cite{{Gaisser}, {Engel2}, {Kope}} 
in efforts to identify contributions that could influence the determination of 
the pp total cross section. The  measured values of   $\sigma_{\rm p-air}^{\rm prod}$
were both apparently too high,   leading  (using models for the 
sigma/slope  correlation) to values of pp cross section then considered too large. 
The Akeno value at $\sqrt{s}=24.54$ TeV is  550 $\pm$ 72 mb , and the 
Fly's Eye measurement at $\sqrt{s}= 30.0$  TeV  is  530 $\pm$ 66 mb .
As seen in Fig. \ref{CR_data-fig}  our calculation also considers  these   
values of production  as  too high. A  critical analysis of the 
interpretation of the experiments \cite{Engel2} showed that the reported values 
for  $\sigma_{\rm p-air}^{\rm prod}$  should be reduced. 
Actually,  a later measurement \cite{Knurenko} 
of the Yakutsk Array experiment obtained  a   comparatively lower value  
525 $\pm$ 52 mb  at 30.65 GeV  that  is closer to our prediction of 509 mb.  
 Contributions due to processes of excitation of nucleon isobars 
\cite{Gaisser}, that were estimated as being at about 3 \%, are not considered in 
other calculations \cite{Engel2,Kope}. 
These measurements and analyses in the 30 TeV region are an example of 
difficulties in the interpretation of EAS data. 

\begin{figure}[b]
\includegraphics[width=8cm]{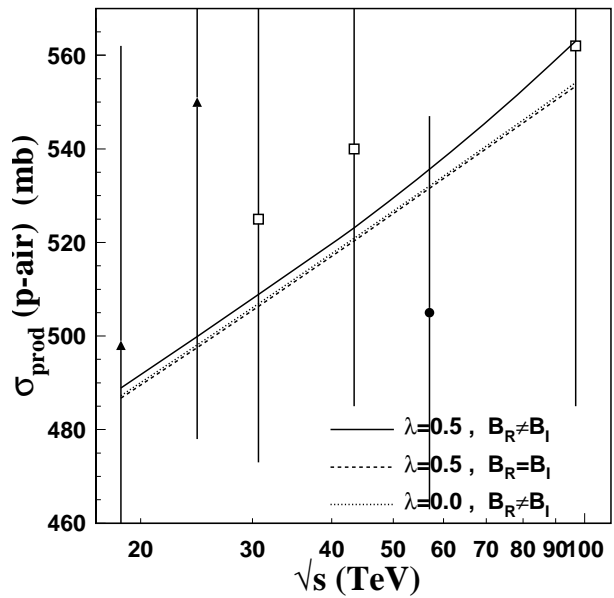}  
\caption{ Effects of the values of the parameter $\protect\lambda$ of the
Good-Walker formalism with intermediate states and of the difference of
values between imaginary and real slopes in Glauber calculation. The solid
line represents the  calculation  with $%
\protect\lambda=0.5$ . The dashed and dotted lines, very close to each
other, represent modified calculations putting $\protect\lambda=0$, in
dotted line, and putting $B_R=B_I$, in dashed line. Some data points are
shown together to help the information on the magnitude of the effects. }
\label{effects-fig}  
\end{figure}

Fig. \ref{effects-fig} shows the influences of the difference of values $B_R
\neq B_I$ and of the quantity $\lambda$ that represents the presence of
diffractive intermediate states, which is tested with values 0 and 0.5 
\cite{Engel,Auger}.  As we see, the effects do not appear as large in the plots,
increase with the energy, and may become more important as experimental
errors and oscillations decrease.
The value $\lambda=0.5$  is assumed to represent the  measurement of 
$\sigma^{\rm SD}/\sigma{\rm inel}$ from ISR. This value could be updated 
with LHC   measurements.  

Table \ref{effects-tab} shows comparative numbers for several cases at the
energy 57 TeV, where we see that the effects on values of the p-air cross
section are under 1 percent. In the $B_R$ case the weak influence is due to
the small $\rho$ value.

\begin{table}[ptb]
\caption{Influences of the quantities $\protect\lambda$ and $B_R$ in Glauber
calculations of $\protect\sigma_{\mathrm{p-air}}$ at $\protect\sqrt{s}= 57$
TeV. The input parameters are $\protect\sigma=140.66$ mb, $B_I=25.33 ~
\nobreak\,\mbox{GeV}^{-2}$ , $B_R=39.80 ~\nobreak\,\mbox{GeV}^{-2}$ and $%
\protect\rho=0.132$ . Some data points are included to provide a scale for
the importance of the effects in comparison to experimental errors. The
effects increase with the energy, and may become important as experimental
errors decrease. \protect\vspace{0.1cm} }
\label{effects-tab}
\tabcolsep=0.009cm 
\begin{tabular}{cccc}
\hline
$\lambda$ & $B_I$ & $B_R$ & $\sigma_{\mathrm{p-air}}^{\mathrm{prod}}$ \\ 
\hline
0.5 ~ & ~ 25.329 ~ & ~ 39.796 ~ & ~ 539.225 \\ 
0.5 ~ & ~ 25.329 ~ & ~25.329 ~ & ~ 536.617 \\ 
0.0 ~ & ~ 25.329 ~ & ~ 39.796 ~ & ~ 537.547 \\ 
0.0 ~ & ~ 25.329~ & ~ 25.329 ~ & ~ 537.333 \\ \hline
\end{tabular}
\end{table}

The   confrontation of our calculation  with data at high energies
does not indicate  the need of contributions beyond the standard Glauber 
calculation. However, the EAS data are  not regular and have large error bars, 
due to uncertainties in the extraction  of   values for  
$\sigma^{\rm prod}_{\rm p-air} $. 
Improvement in the quality of future data may indicate   influence of processes 
  occurring  in intermediate states of the p-air collision, as  nucleon excitations, 
correlations, shadowing.
A particular  example is given by the recent AUGER measurement 
at 57  TeV , that seems a bit too low with respect to the general trend of 
the data, and  has been published  with large error bars.

In the low energy region, the data of the ARGO YBJ collaboration  
 \cite{Aielli}  there may be 
   a regular deviation  of our calculations. It may be that same effects that 
are   not observable at 100 TeV  may become  important   
in this range. Anyhow, the discrepancies are not large, 
amounting to a maximum of 10\% : at $\sqrt{s}=0.0865$ TeV 
the ARGO YBJ experiment gives  $ \sigma_{\rm p-air}^{\rm prod}= 272 \pm 15.8 $ mb , 
while the  theory  gives 307.21 mb. On the contrary, at 
$\sqrt{s}= 0.031$ TeV   the Kaskade experiment \cite{Mielke} and the theoretical value 
coincide very well (at $281 \pm 8.5$  and 286 mb respectively).  
  
In general, there seems  to be more room for improvement in the 
measurements than in our theoretical calculation, and we  believe that 
our pp input together with the basic Glauber calculation have 
  successfully passed the test in the comparison with EAS data.


\section{ Geometric View and Asymptotic Approach  \label{asymptotic-section} }

\begin{figure*}[b]
\includegraphics[width=8cm]{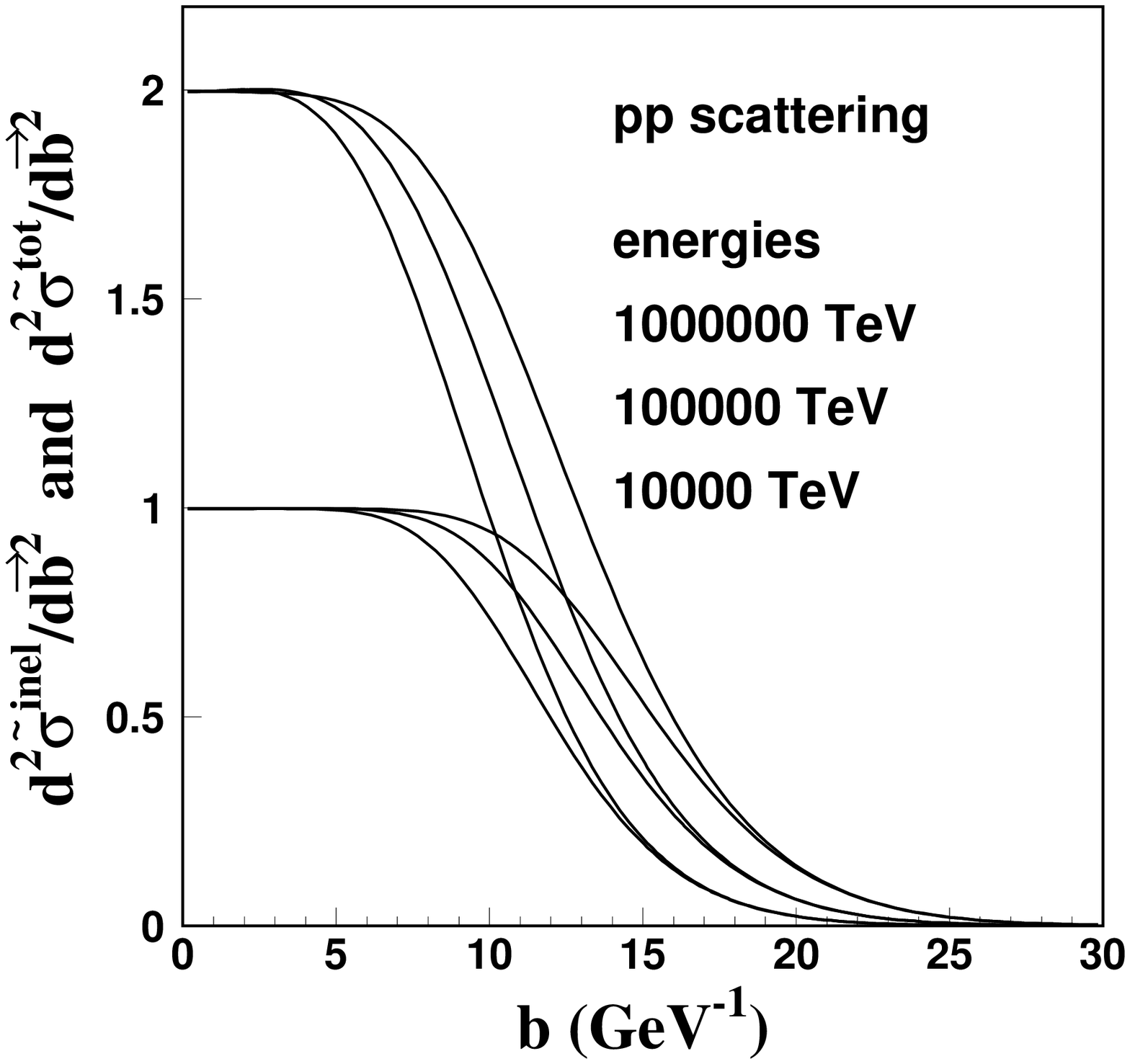} 
\includegraphics[width=8cm]{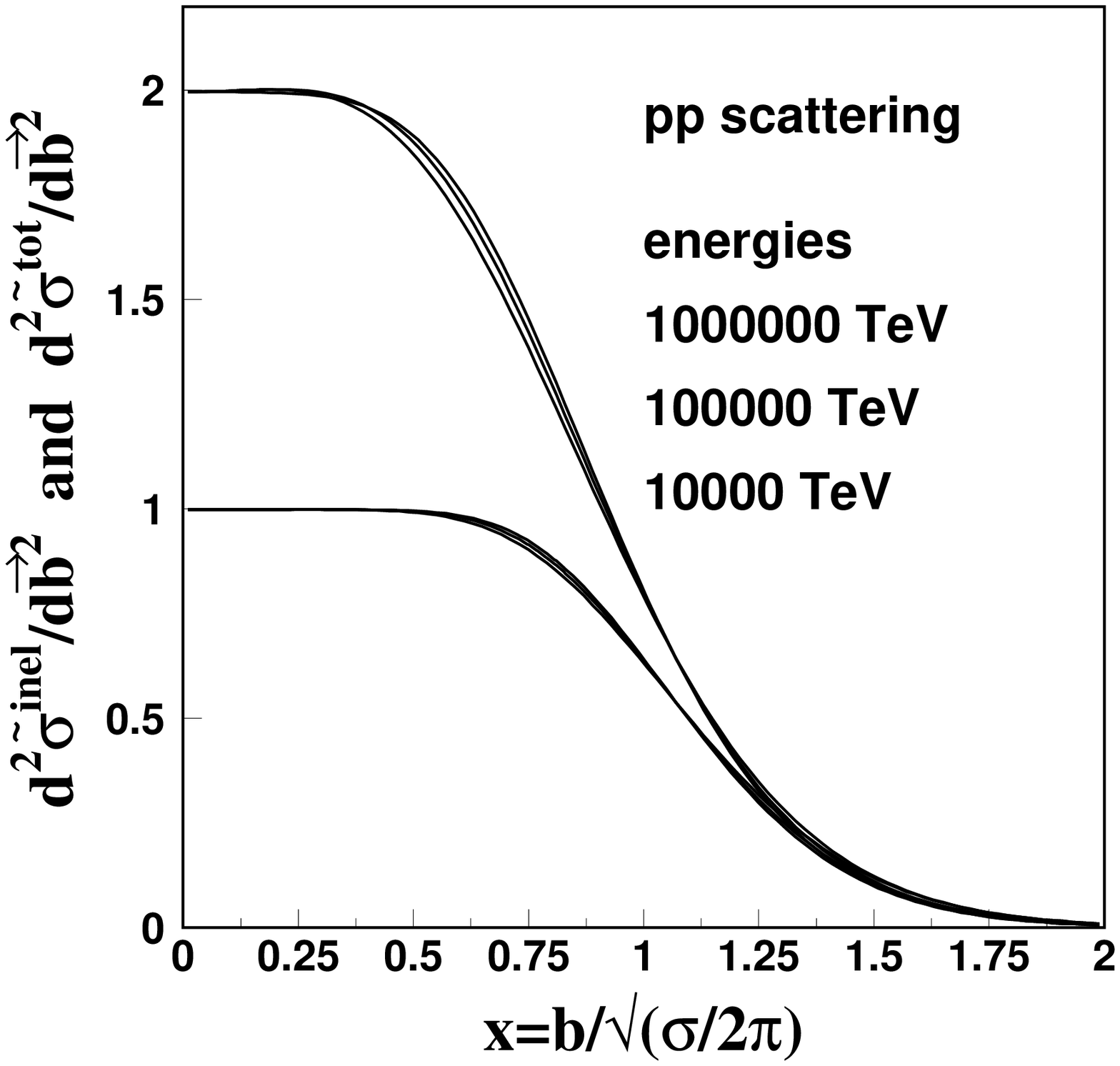}
\caption{Dimensionless differential $b$-space cross sections for total and inelastic
pp interactions. The plotted energies are $10^4$, $10^5$   and $10^6$ TeV. 
In the second part of the figure, the cross sections are plotted against the 
scaled variable $x$, showing universal behaviour, as explained in the text.  }
\label{scaling-fig}
\end{figure*}

An important feature of our pp scattering amplitude is its large-$b$
behaviour.  Writing the $b$ integrated cross section as 
\begin{equation}
\sigma (s) =\int d^{2}\vec{b}\frac{d\sigma }{d^{2}\vec{b%
}},
\end{equation}%
we observe that the integrand, $d\sigma /d^{2}\vec{b},$ as function of $b$
present a long range tail, rather than a sharp cut-off, that is the
characteristic of a black disk model\cite{KEK-to-be}. This behaviour
survives at asymptotic energies, presenting a scaling property as shown in
Fig. \ref{scaling-fig}. In the left side of this figure we show $d\sigma
^{\rm tot}_{\rm pp}/d^{2}\vec{b}$ as function of $b$ for three different energies.  When
these curves are plotted as function of scaled variable $x=b/\sqrt{\sigma
\left( \sqrt{s}\right) /2\pi }$, three curves almost degenerate to a unique
curve as shown in the right side of this figure. Such a property is known as
"geometrical scaling law", advocated by J. Dias de Deus, a long time ago.  
\cite{Geometrial scaling}.    

To make clear how this geometrical scaling nature affects in pp$\ $and pA
cross sections, let us first summarize the simplified Glauber picture below.
When we write the elastic pp scattering amplitude as the form Eq.(\ref%
{iT-eikonal}), 
\begin{equation}
-i~\widehat{T}_{\mathrm{pN}}(s,\vec{b})=1-e^{i\chi \left( s,\vec{b}\right)
}~,
\end{equation}%
the last term is essentially the $S-$matrix in $b$ space. For high energies, 
$b$ represents essentially the angular momentum, so that $\chi $ is (a twice
of) the phase shift. In the presence of inelastic channels, $\chi $ becomes
complex, $\chi =\chi _{R}+i\chi _{I},$ and we can define the impact
parameter representation of partial cross sections in terms of these
functions as 
\begin{align}
\frac{d^{2}\sigma _{pp}^{\mathrm{el}}}{d^{2}\vec{b}}& =1-2\cos \chi
_{R}e^{-\chi _{I}}+e^{-2\chi _{I}}, \\
\frac{d^{2}\sigma _{pp}^{\mathrm{inel}}}{d^{2}\vec{b}}& =1-e^{-2\chi _{I}},
\\
\frac{d^{2}\sigma _{pp}^{\mathrm{tot}}}{d^{2}\vec{b}}& =2\left( 1-\cos \chi
_{R}e^{-\chi _{I}}\right) .
\end{align}%
At high energies, for the calculation of total and integrated cross
sections, we can safely take $\chi _{R}\rightarrow 0$, so that 
\begin{align}
\sigma _{pp}^{\mathrm{el}}(s)& \rightarrow \int d^{2}\vec{b}\ \left(
1-e^{-\chi _{I}}\right) ^{2},  \label{sigpp-el} \\
\sigma _{pp}^{\mathrm{inel}}(s)& \rightarrow \int d^{2}\vec{b}\ \left(
1-e^{-2\chi _{I}}\right) ,  \label{sigpp-inel} \\
\sigma _{pp}^{\mathrm{tot}}(s)& \rightarrow 2\int d^{2}\vec{b}\ \left(
1-e^{-\chi _{I}}\right) .  \label{sigpp-tot}
\end{align}%
The Glauber approximation consists in writing the pA S-matrix as a simple
product of independent scattering centers inside the nucleus,%
\begin{equation}
e^{i\chi _{pA}}\simeq \left\langle \prod_{j=1}^{A}e^{i\chi
_{pN_{j}}}\right\rangle 
\end{equation}%
where $\left\langle {}\right\rangle $ denotes the average over all nucleon
states inside the nucleus and the product $\prod_{i}$ is taken over the
nucleons $N_{j}$. Thus, the pA scattering amplitude is 
\begin{align}
-i\widehat{T}_{\mathrm{pA}}(\vec{b})& =1-e^{i\chi _{pA}} \\
& \simeq 1-\left\langle \prod_{j=1}^{A}e^{i\chi _{pN_{j}}}\right\rangle  
\nonumber \\
& =1-\left\langle \prod_{j=1}^{A}\left( 1+i\widehat{T}_{\mathrm{pN}}(\vec{b}%
)\right) \right\rangle ,  \nonumber
\end{align}%
that leads to equations 
of last section. Glauber approach gives essentially 
\begin{eqnarray}
&&\frac{1}{2}\frac{d^{2}\sigma _{pA}^{\mathrm{tot}}}{d^{2}\vec{b}}(s,\vec{b})
\\
&=&\left\langle 1-\prod_{i=1}^{A}\left( 1-\frac{1}{2}\frac{d^{2}\sigma
_{pp}^{\mathrm{tot}}}{d^{2}\vec{b_{i}}}(s,\vec{b}-\vec{b}_{i})\right)
\right\rangle ,  \nonumber
\end{eqnarray}%
and 
\begin{eqnarray}
&&\frac{d^{2}\sigma _{pA}^{\mathrm{el}}}{d^{2}\vec{b}}(s,\vec{b})  \nonumber
\\
&=&\left\langle \bigg[1-\prod_{i=1}^{A}(1-\frac{d^{2}\sigma _{pp}^{\mathrm{%
tot}}}{d^{2}\vec{b_{i}}}(s,\vec{b}-\vec{b}_{i}))\bigg]^{2}\right\rangle .
\end{eqnarray}%
At extremely high energies, $\sigma _{pp}^{\mathrm{tot}}$ may become much
larger than the geometrical cross section of the target nucleus, $\sigma
_{A}^{\mathrm{geo}}\equiv \pi R_{A}^{2},$ where $R_{A}$ is the nuclear
radius. In such a situation we may neglect the variation in position of each
nucleon $\left( \vec{b}_{i}\sim 0\right) $, and we can approximate 
\begin{equation}
\frac{1}{2}\frac{d^{2}\sigma _{pA}^{\mathrm{tot}}}{d^{2}\vec{b}}(s,\vec{b}%
)\simeq 1-\left( 1-\frac{1}{2}\frac{d^{2}\sigma _{pp}^{\mathrm{tot}}}{d^{2}%
\vec{b}}\left( s,\vec{b}\right) \right) ^{A}~,  \label{Glauber-Tot}
\end{equation}%
and 
\begin{equation}
\frac{d^{2}\sigma _{pA}^{\mathrm{el}}}{d^{2}\vec{b}}\left( s,\vec{b}\right)
\simeq \left[ 1-\left( 1-\frac{d^{2}\sigma _{pp}^{\mathrm{tot}}}{d^{2}\vec{b}%
}\left( s,\vec{b}\right) \right) ^{A}\right] ^{2}.  \label{Glauber-ela}
\end{equation}%
Such situation can occur in our case only for $\sqrt{s}\gg 10^{12}$ TeV,
much larger than the highest energy observed in cosmic ray experiments.

Now, as shown in Fig, (\ref{scaling-fig}), our amplitudes lead to an
approximate geometric scaling law for very large energies,%
\begin{equation}
\frac{1}{2}\frac{d\sigma _{pp}^{\mathrm{tot}}}{d^{2}\vec{b}}(s,\vec{b}%
)~\rightarrow ~\zeta \left( x\right) ~,
\end{equation}%
where $\zeta $ is an unversal function independent of $\sqrt{s}$ and%
\begin{equation}
x\equiv \frac{b}{b_{eff}\left( \sqrt{s}\right) }~,
\end{equation}%
with $b_{eff}(\sqrt{s})\sim \sigma \left( \sqrt{s}\right) $. The total pp
cross section then becomes 
\begin{equation}
\sigma _{pp}^{\mathrm{tot}}\left( s\right) \rightarrow ~4\pi
b_{eff}^{2}\left( \sqrt{s}\right) \int_{0}^{\infty }x~~\zeta \left( x\right)
\ dx~.  \label{sigppp-tot-scale}
\end{equation}%
If we introduce another function 
\begin{equation}
\xi (x)~=~1-\left[ 1-\zeta (x)\right] ^{2},  \label{gzi-zeta}
\end{equation}%
to write the inelastic cross section as 
\begin{equation}
\sigma _{pp}^{\mathrm{inel}}\left( s\right) ~\rightarrow ~2\pi
b_{eff}^{2}\left( \sqrt{s}\right) \int_{0}^{\infty }x~\xi \left( x\right) \
dx~,  \label{sigpp-inel-scale}
\end{equation}%
where we have used Eqs. (\ref{sigpp-inel}, \ref{sigpp-tot}). From Eqs.(\ref%
{sigppp-tot-scale}, \ref{sigpp-inel-scale}), we obtain 
\begin{equation}
\frac{\sigma _{pp}^{\mathrm{inel}}\left( s\right) }{\sigma _{pp}^{\mathrm{tot%
}}\left( s\right) }\rightarrow \frac{\int_{0}^{\infty }x~\xi (x)~dx}{%
2\int_{0}^{\infty }x~\zeta \left( x\right) ~dx}~=~\mathrm{const}.
\label{Constraint}
\end{equation}%
As shown in Fig.(\ref{scaling-fig}), $\zeta $ and $\xi $ are functions having a
common property,%
\begin{equation}
\zeta (x)~,~\xi (x)\rightarrow \left\{ 
\begin{array}{c}
1, \\ 
0,%
\end{array}%
\begin{array}{c}
x\,\rightarrow 0 \\ 
x\rightarrow \infty 
\end{array}%
\right. ,
\end{equation}%
When we have the case of a sharp cut-off of $\zeta $ as in a black disk 
\begin{equation}
\zeta \left( x\right) =\theta \left( 1-x\right) ,  \label{zeta=theta}
\end{equation}%
then $\xi (x)$ becomes identical with $\zeta (x)$ , and we have the ratio 
\begin{equation}
\lim_{{s}\rightarrow \infty }\frac{\sigma _{pp}^{\mathrm{inel}}\left(
s\right) }{\sigma _{pp}^{\mathrm{tot}}\left( s\right) }=\frac{1}{2}~,
\end{equation}%
that is a well known result for a black disk.

Generally, $\zeta (x)$ is not a sharp-cut theta function as in Eq. (\ref%
{zeta=theta}) but stays unity up to a certain value of $x$ (that is $x=1$, $%
b=b_{eff}(\sqrt{s})$ ), then monotonically decreases to zero with a tail
form.  Let us write then 
\begin{equation}
\zeta \left( x\right) =\left\{ 
\begin{array}{c}
1, \\ 
\Phi \left( x\right) ,%
\end{array}%
\begin{array}{c}
x\,\leq 1 \\ 
x>1%
\end{array}%
\right. ~,
\end{equation}%
where $\Phi \left( x\right) $ is a positive and monotonically decreasing
function with $\Phi \left( 1\right) =1.\ $

Let us now turn to the pA case. From Eqs. (\ref{Glauber-Tot}, \ref%
{Glauber-ela}), we have 
\begin{equation}
\frac{1}{2}\sigma _{pA}^{\mathrm{tot}}(s)=~2\pi b_{eff}^{2}\left( \sqrt{s}%
\right) \int_{0}^{\infty }x~dx\left[ 1-\left( 1-\zeta \left( x\right)
\right) ^{A}\right] ,
\end{equation}%
and%
\begin{equation}
\sigma _{pA}^{\mathrm{el}}(s)=~2\pi b_{eff}^{2}\left( \sqrt{s}\right)
\int_{0}^{\infty }x~dx\left[ 1-\left( 1-\zeta \left( x\right) \right) ^{A}%
\right] ^{2}~,
\end{equation}%
so that, taking the difference $\sigma _{pA}^{\mathrm{tot}}-\sigma _{pA}^{%
\mathrm{el}}$ , 
\begin{align}
& \sigma _{pA}^{\mathrm{inel}}(s)=~2\pi b_{eff}^{2}\left( \sqrt{s}\right)
\int_{0}^{\infty }x~dx\left[ 1-\left( 1-\zeta \left( x\right) \right) ^{2A}%
\right]  \\
& =~2\pi b_{eff}^{2}\left( \sqrt{s}\right) \left( \frac{1}{2}%
+\int_{1}^{\infty }x~dx\left[ 1-\left( 1-\Phi \left( x\right) \right) ^{2A}%
\right] \right) ~.  \nonumber
\end{align}%
Since $0\leq 1-\Phi \leq 1$ for all $x,$ we have $\left( 1-\Phi \right)
^{2A}\leq 1-\Phi ,$ for $A\geq 1.$ Thus we have the inequality 
\begin{equation}
\int_{1}^{\infty }x~dx\left( 1-\left( 1-\Phi (x)\right) ^{2A}\right) \geq
\int_{1}^{\infty }x~dx~\Phi \left( x\right) .
\end{equation}%
From this consideration, we arrive at the conclusion that 
\begin{align}
\frac{\sigma _{pA}^{\mathrm{inel}}}{\sigma _{pp}^{\mathrm{tot}}}(s)&
=\int_{0}^{\infty }x~dx\left[ 1-\left( 1-\zeta \left( x\right) \right) ^{2A}%
\right] /\int_{0}^{\infty }2x~\zeta \left( x\right) \ dx  \nonumber \\
& \geq 1/2~,  \label{RatioGlauber/pp}
\end{align}%
for $\sqrt{s}\rightarrow \infty .\ $ Note that in the black disk case $\Phi
\left( x\right) \equiv 0,$ or equivalently $\zeta \left( x\right) =\theta
\left( 1-x\right) ,\ $we obtain the well-defined limit 
\[
\sigma _{pA}^{\mathrm{inel}}(s)/\sigma _{pp}^{\mathrm{tot}}(s)\rightarrow 
\frac{1}{2}~.
\]%
As a corollary to Eq.(\ref{RatioGlauber/pp}), for two different target
nuclei $A$ and $A^{\prime }$, with   
for $A<A^{\prime }$ and $\zeta \left( x\right) \neq \theta \left( x\right) $ 
we have the inequality 
\begin{equation}
\frac{\sigma _{pA}^{\mathrm{inel}}}{\sigma _{pp}^{\mathrm{tot}}}(s)<\frac{%
\sigma _{pA^{\prime }}^{\mathrm{inel}}}{\sigma _{pp}^{\mathrm{tot}}}(s) ~ . 
\end{equation}%

Naturally Eq.(\ref{RatioGlauber/pp}) is valid also for $A=1$ and in this case 
\begin{equation}
\frac{\sigma _{pp}^{\mathrm{inel}}}{\sigma _{pp}^{\mathrm{tot}}}(s)=\frac{%
1+2\int_{1}^{\infty }x~dx\left[ 1-\left( 1-\Phi \left( x\right) \right) ^{2}%
\right] }{2\left( 1+2\int_{1}^{\infty }x~dx\Phi \left( x\right) \right) }>%
\frac{1}{2},
\end{equation}%
if $\Phi \neq 0.$ We thus see  that the non-black disk nature is intimately related
to the tail property $\Phi \left( x\right) $.

As mentioned before, our phenomenological pp representation does not
correspond to the black disk, and the actual pp ratio is $\sigma _{pp}^{%
\mathrm{inel}}/\sigma _{pp}^{\mathrm{tot}}~\rightarrow ~2/3$ . This
constraints the tail $\Phi$, %
\begin{equation}
\frac{1+2\int_{1}^{\infty }x~dx\left[ 1-\left( 1-\Phi \left( x\right)
\right) ^{2}\right] }{2\left( 1+2\int_{1}^{\infty }x~dx\Phi \left( x\right)
\right) }=\frac{2}{3}.  \label{Constriant2}
\end{equation}%
With this information at hand, we look for an estimate of the value 
 $$ \sigma _{pA}^{\mathrm{inel}}(s)/\sigma _{pp}^{\mathrm{tot}}(s)$$ 
 using a tail form proper for the realistic pp amplitudes.

As a simple choice, considering that the stochastic vacuum model predicts
the tail as that of Yukawa behaviour for large $b$, we take 
\begin{equation}
\zeta \left( x\right) =\left\{ 
\begin{array}{c}
1, \\ 
\exp (-\alpha (x-1))/x,%
\end{array}%
\begin{array}{c}
x\,\leq 1 \\ 
x>1%
\end{array}%
\right. ,  \label{zeta}
\end{equation}%
where $\alpha $ is a parameter to be determined using Eq.(\ref{Constriant2})
In this case, we have%
\begin{equation}
\frac{1}{2}\sigma _{\mathrm{pp}}^{\mathrm{tot}}=~2\pi b_{eff}^{2}\left( 
\sqrt{s}\right) \left( \frac{1}{2}+\frac{1}{\alpha }\right) ,
\end{equation}%
and 
\begin{equation}
\xi (x)=\left\{ 
\begin{array}{c}
1, \\ 
2e^{-\alpha \left( x-1\right) }/x-e^{-2\alpha \left( x-1\right) }/x^{2},%
\end{array}%
\begin{array}{c}
x\,\leq 1 \\ 
x>1%
\end{array}%
\right. 
\end{equation}%
to obtain 
\begin{equation}
\sigma _{\mathrm{pp}}^{\mathrm{inel}}=~2\pi b_{eff}^{2}\left( \sqrt{s}%
\right) \left( \frac{1}{2}+\frac{2}{\alpha }-\int_{0}^{\infty }\frac{%
e^{-2\alpha x}}{x+1}dx\right) 
\end{equation}%
The constraint for $\alpha $ from Eq.(\ref{Constriant2}) becomes%
\begin{equation}
4\left( \frac{1}{2}+\frac{1}{\alpha }\right) =3\left( \frac{1}{2}+\frac{2}{%
\alpha }-\int_{0}^{\infty }\frac{e^{-2\alpha x}}{x+1}dx\right) ~,
\end{equation}%
%
%
%
leading to 
\begin{equation}
\alpha \simeq 1.61073~.
\end{equation}%
With this, for $A=15,$ for example, we obtain 
\begin{eqnarray}
~\sigma _{\mathrm{pA}}^{\mathrm{inel}} &=&~2\pi b_{eff}^{2}\left( \sqrt{s}%
\right) \left( 1+\int_{1}^{\infty }x~dx\left( 1-\frac{e^{-\alpha (x-1)}}{x}%
\right) ^{2A}\right)   \nonumber \\
&\simeq &~2\pi b_{eff}^{2}\left( \sqrt{s}\right) \times 2.30764~,   
\end{eqnarray}%
giving%
\begin{equation}
\left. \frac{\sigma _{\mathrm{pA}}^{\mathrm{inel}}}{\sigma _{\mathrm{pp}}^{%
\mathrm{tot}}}\right\vert _{\rm Yukawa}\simeq 1.1858~.  \label{ratio3}
\end{equation}

This value depends sensitively on the choice of the tail function $\Phi .$
The slower the decay of the tail, the bigger the ratio becomes. If we choose 
$\Phi $ a pure exponential, 
\begin{equation}
\Phi =e^{-\alpha \left( x-1\right) },
\end{equation}%
which is more longer tail than Yukawa type, then using the same procedure to
get $\alpha \simeq $ $2.\,\allowbreak 158\,3$ and the corresponding value of
the ratio becomes%
\begin{equation}
\left. \frac{\sigma _{\mathrm{pA}}^{\mathrm{inel}}}{\sigma _{\mathrm{pp}}^{%
\mathrm{tot}}}\right\vert _{\rm Exponential}\simeq 1.798\ .  \label{ratio4}
\end{equation}

\begin{figure*}[b]
\includegraphics[width=8cm]{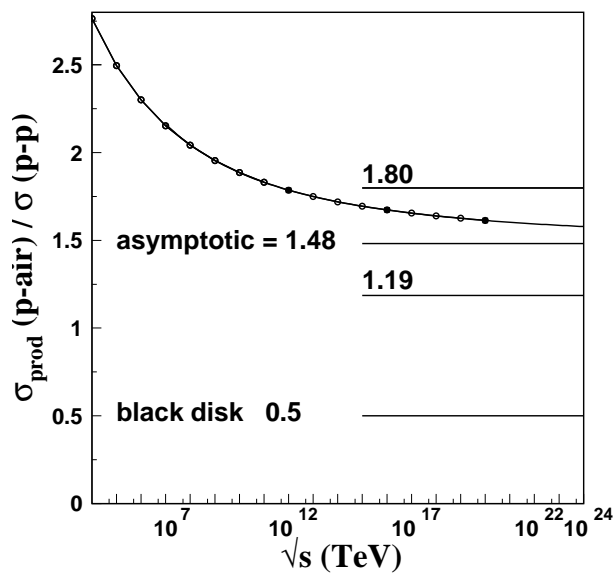} 
\caption{ Ratio of p-air and pp cross sections at ultra-high energies.
Calculations are marked with dots and connected with a continuous line. The
dashed line is given analytically by the fraction of $\log ^{2}$ forms for $%
\protect\sigma _{\mathrm{p-air}}^{\mathrm{inel}}(s)$ and $\protect\sigma _{%
\mathrm{pp}}^{\mathrm{tot}}(s)$, given in the text. It gives good
representation of the points for energies above $10^{6}$ TeV and tends to
the asymptotic limit 1.48, as explained in the text ~ .}
\label{asymptotic_ratio-fig}
\end{figure*}

These values of ratio for different tails can be compared with the energy
dependence of the ratio shown in Fig. \ref{asymptotic_ratio-fig} where we
plotted the ratio calculated directly by integrating our cross sections
numerically for extremely large $\sqrt{s}$ values up to $\sqrt{s}=10^{20}$
TeV. We note that the values are still decreasing, but approaches to a value
between those given in Eqs.(\ref{ratio3}) and (\ref{ratio4}) .

As we see from this figure, the asymptotic value is only attained only for
really large $\sqrt{s},$ say $\sqrt{s}\gg 10^{20} $ TeV.  Numerical integration
of the cross section at such values of $\sqrt{s}$ is not trivial due to the
huge cancellations, but just to see the tendency, we use the values of $%
\sigma _{\mathrm{p-air}}^{\mathrm{inel}}$ at $10^{12}$, $10^{16}$ and $%
10^{20}$ to obtain the extrapolation form 
\begin{equation}
\sigma _{\mathrm{p-air}}^{\mathrm{inel}}(s)=490.883+19.7119\log \sqrt{s}%
+1.8178\log ^{2}\sqrt{s}~.  \label{large_s}
\end{equation}%
Dividing this function by the $\log ^{2}$ form of the pp total cross section
in Eq. (\ref{sig-eq}), we obtain the dashed line shown in the figure. We see
that the representation of the ratio looks very good above $10^{6}$ TeV. In
this parametrization the predicted asymptotic limit is $%
1.8178/1.2273=1.4811~.$ We would obtain somewhat different limit, had we
taken a different set of three energies to construct the form in Eq. (\ref%
{large_s}), but the result would remain in the interval 1.4 - 1.5 . The slow
convergence of the ratio towards a finite limit at high energies is an
important fact.

The uncertainties given in Eqs.(\ref{ratio3}) or (\ref{ratio4}) are due to
the form of ansatz, $\zeta $. A sharp transition like Eq.(\ref{zeta}) at $%
x=1 $ is not realistic to our amplitude. However, it is interesting to note
that the extrapolated numerical value is in between the values of Eq. (\ref%
{ratio3}) and (\ref{ratio4}), that was determined using as input the 2/3
ratio of inelastic to total pp cross sections and assumption of the
Yukawa-like or Exponential tail in the $b$ dependence of the pp amplitudes.

Eq. (\ref{large_s}) gives a proper representation of $\sigma _{\mathrm{p-air}%
}^{\mathrm{inel}}(s)$ to be used only for energies higher than $\sqrt{s}%
\approx 10^{6}$ TeV. Nonetheless, when used at the highest CR experimental
energy $\sqrt{s}=96.85$ it gives a value just 10\%  larger than the correct
one: thus not too bad. 

On the other hand, the form given for $\sigma _{\mathrm{p-air}}^{\mathrm{inel%
}}(s)$ in Eq. (\ref{curve_data}) is based on the three points $\sqrt{s}$ =
96.85 , $10^{3}$ and $10^{4}$ TeV, and gives very good representation of the
exact values from 10 GeV to $10^{6}$ TeV. However, this form is not adequate
for the asymptotic limit.

The good coherence of different evaluations of these finite asymptotic
ratios is very interesting. They point out to what can be expected for CR
experiments at ultra high energies.


\section{Final Remarks and Comments}

The amplitudes that we have constructed to describe accurately the pp
elastic differential cross sections at energies from 20 GeV to 8 TeV are
used in Glauber formalism to evaluate the p-air production cross section
obtained in EAS/CR experiments. Our prediction for the whole energy interval
from 10 GeV to 100 TeV of p-air production cross section is shown in Fig. %
\ref{CR_data-fig}.

The comparison of our results with data shows good agreement, confirming
that the extrapolation of the input quantities extrapolated to energies
higher by one order of magnitude is consistent. From this we are confident
that our representation of pp scattering amplitudes can be used for higher
CR data.

The calculations with Glauber approach depend crucially on the input values
of $\sigma_{\mathrm{pp}}^{\mathrm{tot}}(s)$ and $B_I(s)$, and thus the
results obtained for the high energies of the CR experiments 
are important tests of the energy dependences that we propose for these
quantities, given in Eqs. (\ref{sig-eq}, \ref{BI-eq}). It is particularly
remarkable that the $\log^2$ dependence that we propose for $B_I(s)$
predicts higher values for the extrapolated values of this quantity, and the
data seem to be consistent with this. Thus at 57 TeV we have $B_I= 25 ~%
\nobreak\,\mbox{GeV}^{-2}$ , value that is higher than the usual obtained,
for example from Donnachie-Landshoff or Regge form. The comparison with CR
data helps to test such alternatives.

The extraction of fundamental information on the energy dependence of pp
total cross section from CR/EAS measurements depends on this point. Thus our
prediction for pp cross section at 57 TeV is of 140.7 mb. In the
experimental paper \cite{Auger}, where the measured value for $\sigma_{%
\mathrm{p-air}}^{\mathrm{prod}}$ is below our calculation (see Fig. \ref%
{CR_data-fig}), and other theoretical models for $\sigma(s)$ and $B_I(s)$
are used, the reported value for $\sigma$ is $133 \pm 29$ mb. Hopefully this
important question will be investigated in future  measurements with cosmic
rays.

An important point of our description of differential elastic cross section
is that we keep full respect for the real part of the scattering amplitude.
The real part is crucial for large $|t|$ but often neglected in the forward
region due to the small value of the $\rho$ parameter. We stress that the
neglect of the proper $B_R$ value affects the determination of pp total
cross section. We take this into account in Glauber calculation of p-air
processes. The influence is not large ($\sim 1\% $ for the total cross section
at 57 TeV), but increases with the energy. We have shown in Fig. \ref%
{effects-fig} and in Table \ref{effects-tab}  that the effects of the
condition $B_R > B_I$ and of the presence  of intermediate diffractive
states (parameter $\lambda$) in Good-Walker \cite{Good_Walker} approach are
of similar magnitudes.

From our representation of the scattering amplitudes we can calculate the
asymptotic values of quantities that approach finite values at high
energies. These values are important for the geometric interpretation of the
dynamics, as can be studied in the representation of the impact parameter $b$%
. For example, the behaviour of the ratios $\sigma_{\mathrm{pp}}^{\mathrm{tot%
}}/B_I$ and $\sigma_{\mathrm{pp}}^{\mathrm{tot}}/B_R$ are connected with
integrated elastic pp cross sections and thus with the rate of inelastic
proccesses at high energies in the pp system. Our result shows that the
ratio, $\sigma_{\mathrm{pp}}^{\mathrm{inel}}/\sigma_{\mathrm{pp}}^{\mathrm{%
tot}} => \approx 2/3 $ at very high energies.

To acquire a better feeling about the regularity of the energy dependence of
the data and its representation by the theoretical calculation, we present
in Fig. \ref{ratio_sigmas-fig} results on the ratio between p-air and pp
cross sections. The figure shows that this ratio has the important property
of approaching a finite value for infinite energy. This information if of
fundamental importance for the understanding of the geometric nature of the
pp interaction and its energy dependence. The question is investigated in
Sec. \ref{data-section}  within the Glauber formalism. We show that this ratio is intimately
related with the ratio sigma(pp inelastic)/sigma(pp total) and with  the
behaviour of the eikonal functions for large b.

The important question of the energy dependence of the ratio of p-air to pp
cross sections is studied in a direct way, using properties of the $b$
dependence of pp interaction at high energies. We show that the Yukawa-like
behaviour of the interaction range, inspired in the stochastic vacuum model,
explains quantitatively with high accuracy the value of the asymptotic limit
of the ratio $\sigma_{\mathrm{p-air}}^{\mathrm{inel}}/\sigma_{\mathrm{pp}}^{%
\mathrm{tot}} $.

This is what we have, considering that the nucleons are the scattering
centers in Glauber framework. Of course, for a ultra-high energy domain,
where the pp cross section overcomes the geometric cross section of a target
nucleus, the Glauber approach itself may be questionable. In the Glauber
approach of pA cross section, the scattering centers inside the target are
nucleons, with a fixed distribution determined by the nuclear wave function.
However, at the energies where the interaction size of pp becomes large
enough so that their superposition becomes not negligible, the scattering
centers are rather partons and not nucleons. Then the energy dependence of
pA cross section can become drastically different \cite{licinio} . Here we
have an open question. Further theoretical investigations of microscopic
structures leading to the asymptotic behavior in p-air cross cross sections
will be very interesting.

\begin{acknowledgments}
The authors wish to thank the Brazilian agencies CNPq, PRONEX and FAPERJ for 
financial support.  
\end{acknowledgments}

\end{document}